\documentclass[twocolumn,conference]{IEEEtran}
\usepackage[T1]{fontenc}
\usepackage[latin9]{inputenc}
\usepackage{color}
\usepackage{bm}
\usepackage{amsmath}
\usepackage{amsthm}
\usepackage{amssymb}
\usepackage{stackrel}
\usepackage{graphicx}

\makeatletter

\theoremstyle{plain}
\newtheorem{thm}{\protect\theoremname}
\theoremstyle{plain}
\newtheorem{lem}[thm]{\protect\lemmaname}

\usepackage[caption=false,font=footnotesize]{subfig}
\usepackage{algorithmic}
\usepackage{algorithm}

\usepackage{bm}
\usepackage{cite}
\usepackage[english]{babel}
\usepackage[T1]{fontenc}

\IEEEoverridecommandlockouts

\makeatother

\providecommand{\lemmaname}{Lemma}
\providecommand{\theoremname}{Theorem}

\begin{document}
\title{Average Age of Changed Information in the Internet of Things}
\author{\IEEEauthorblockN{Wenrui~Lin\IEEEauthorrefmark{1}\IEEEauthorrefmark{2}, Xijun~Wang\IEEEauthorrefmark{1}\IEEEauthorrefmark{2},
Chao Xu\IEEEauthorrefmark{3}, Xinghua Sun\IEEEauthorrefmark{1},
and Xiang Chen\IEEEauthorrefmark{4} }\IEEEauthorblockA{\IEEEauthorrefmark{1}School of Electronics and Communication Engineering,
Sun Yat-sen University, Guangzhou, 510006, China\\
\IEEEauthorrefmark{2}Key Laboratory of Wireless Sensor Network \&
Communication, \\
Shanghai Institute of Microsystem and Information Technology, \\
Chinese Academy of Sciences, 865 Changning Road, Shanghai 200050 China\\
\IEEEauthorrefmark{3}School of Information Engineering, Northwest
A\&F University, Yangling, Shaanxi, China\\
\IEEEauthorrefmark{4}School of Electronics and Information Technology,
Sun Yat-sen University, Guangzhou, 510006, China\\
Email: linwr7@mail2.sysu.edu.cn, wangxijun@mail.sysu.edu.cn, cxu@nwafu.edu.cn,
\\
sunxinghua@mail.sysu.edu.cn, chenxiang@mail.sysu.edu.cn}\thanks{This work was supported in part by the National Natural Science Foundation
of China (61701372), by the Research Fund of the Key Laboratory of
Wireless Sensor Network \& Communication (Shanghai Institute of Microsystem
and Information Technology, Chinese Academy of Sciences) under grant
20190912, by Fundamental Research Funds for the Central Universities
under 19lgpy79 and 19lgpy77, by Talents Special Foundation of Northwest
A\&F University (Z111021801), by Guangdong Basic and Applied Basic
Research Foundation (2019A1515011906), by State's Key Project of Research
and Development Plan (No.2017YFE0121300-6), and by Guangdong Provincial
Special Fund For Modern Agriculture Industry Technology Innovation
Teams (No.2019KJ122).}}
\maketitle
\begin{abstract}
The freshness of status updates is imperative in mission-critical
Internet of things (IoT) applications. Recently, Age of Information
(AoI) has been proposed to measure the freshness of updates at the
receiver. However, AoI only characterizes the freshness over time,
but ignores the freshness in the content. In this paper, we introduce
a new performance metric, \emph{Age of Changed Information} (AoCI),
which captures both the passage of time and the change of information
content. Also, we examine the AoCI in a time-slotted status update
system, where a sensor samples the physical process and transmits
the update packets with a cost. We formulate a Markov Decision Process
(MDP) to find the optimal updating policy that minimizes the weighted
sum of the AoCI and the update cost. Particularly, in a special case
that the physical process is modeled by a two-state discrete time
Markov chain with equal transition probability, we show that the optimal
policy is of threshold type with respect to the AoCI and derive the
closed-form of the threshold. Finally, simulations are conducted to
exhibit the performance of the threshold policy and its superiority
over the zero-wait baseline policy. 
\end{abstract}

\section{Introduction}

With the sharp proliferation of the Internet of Thing (IoT) devices
and the rising need of mission-critical services, timely delivery
of information has become increasingly important in real-time status
update systems \cite{palattellaInternetThings5G2016,schulzLatencyCriticalIoT2017}.
The performance of such systems depends on the freshness of the status
updates received by the destination \cite{liuAgeoptimalTrajectoryPlanning2018,tongUAVEnabledAgeOptimalData2019,xuOptimizingInformationFreshness2019}.
Recently, the age of information (AoI) has been introduced to measure
data freshness from the receiver's perspective \cite{kaulRealtimeStatusHow2012}.
In particular, it is defined as the time elapsed since the generation
of the most recent status update packet received by the destination.
Essentially, AoI jointly characterizes the packet delay and the packet
inter-generation time, which distinguishes AoI from conventional delay
metrics. However, it ignores the content carried by the updates and
the current knowledge of the receiver.

A natural question that arises then is whether it is sufficient to
measure the freshness of updates via AoI only. There have been some
recent efforts to answer this question. In \cite{sunSamplingDataFreshness2018},
the mutual information between the state of the source and the received
updates at the destination was defined as the freshness metric, which
was proved to be a non-negative and non-increasing function of AoI
if the sampling times are independent of the state of the source.
For more general sampling patterns, the AoI is inadequate to reflect
the freshness in information content and hence different metrics have
been proposed in {[}8{]}-{[}10{]}. In \cite{kamEffectiveAgeInformation2018a},
the authors proposed a metric, named sampling age, which is the time
difference between the last ideal sampling time and the first actual
sampling time. The sampling age is monotonically increasing with respect
to estimation error for a Markov source, but the ideal sampling time
is nontrivial to obtain. Age of synchronization (AoS) was proposed
in \cite{zhongTwoFreshnessMetrics2018} to measure the time that the
process being tracked has changed. Particularly, AoS is defined as
the time difference between the current time and the first update
time after the previous synchronization time. Actually, it is implicitly
assumed that the first update after each synchronization contains
new information. The authors in \cite{maatoukAgeIncorrectInformation2019}
proposed age of incorrect information (AoII) as a new metric by combining
time and estimation error penalty functions. As such, the AoII will
increase with time when the receiver stays in an erroneous state.
Note that an estimation error occurs when the current estimate at
the receiver is different from the actual state of the process. Nonetheless,
such an actual state cannot be perceived by the receiver unless the
related update is delivered and hence, exactly depicting the AoII
at the receiver between two successful transmissions is far from being
trivial.\textcolor{red}{}

In this paper, we first introduce a new performance metric, referred
to as \emph{age of changed information} (AoCI), that characterizes
the information freshness via both the passage of time and the change
of information content.  Then, we study the AoCI in a status update
system consisting of a sensor and a destination. In particular, the
sensor monitors the real-time status of a physical process, which
is modeled by a two-state discrete time Markov chain, and transmits
status update packets to the destination through a wireless channel,
which incurs an update cost. We aim to find the optimal updating policy
that minimizes the total average cost, which is the weighted sum of
the AoCI and the update cost. By formulating this problem into a Markov
decision process (MDP), we prove that the optimal updating policy
is a threshold-type policy and further derive the threshold in closed-form
with a special Markov chain model of the physical process. Simulation
results show that the threshold policy can achieve lower total average
cost than the zero-wait policy.

The rest of the paper is organized as follows: Section II presents
the system model and introduces the proposed metric. In Section III,
we provide the MDP formulation of the problem, analyze the switching
structure of the optimal policy, and derive the threshold in closed-form.
Simulation results are presented in Section IV, followed by the conclusion
in Section V.

\section{System Overview \label{sec:System-Overview}}

\subsection{System Model}

We consider a time-slotted status update system which consists of
a sensor and a destination (e.g., a monitor or an actuator). In each
time slot, the sensor could remain idle to save energy. Or it could
generate a status update about the underlying time-varying process
(a.k.a. generate-at-will) and send it to the destination over an unreliable
channel to refresh the destination. Let $a_{t}\in\{0,1\}$ be the
action of the sensor in the $t$-th slot, where $a_{t}=1$ indicates
that the sensor samples and transmits a new update, and $a_{t}=0$,
otherwise. In general, there will be a cost associated with each update.
We let $C_{u}$ denote the cost of an update. Moreover, the transmission
time of each update is assumed to be equal to the duration of one
time slot. Without loss of generality, the slot duration is normalized
to unity. 
\begin{figure}[tp]
\centering

\includegraphics[width=0.45\textwidth]{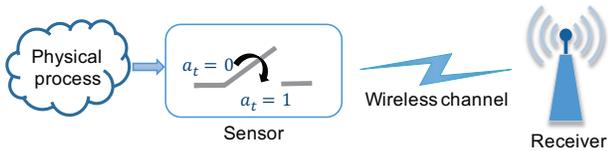}\caption{\label{fig:Transmission-Time}A model of a status update system.}
\end{figure}

Assume that the underlying time-varying physical process is modeled
by a two-state discrete time Markov chain $\{X_{t};t\in\mathbb{N}\}$
with $X_{t}\in\{0,1\}$, where the duration of each state is equal
to the slot length and the transition occurs just prior to the sampling
decision at the beginning of each slot. The one-step state transition
probability matrix is given by
\begin{equation}
\left[\begin{array}{cc}
1-p_{c} & p_{c}\\
p_{c} & 1-p_{c}
\end{array}\right],\label{eq:transition-matrix}
\end{equation}
where $p_{c}\in(0,1)$ is the probability of changing states. 

We assume that channel fading remains constant in each slot but independently
changes over different slots. We also assume that the sensor transmits
an update at a fixed rate and the channel state information is available
only at the destination. As such, the transmission in each time slot
may fail due to outage and the packet loss could be characterized
by a memoryless Bernoulli process. Specifically, let $h_{t}\in\{0,1\}$
denote whether the transmission succeeds or fails, where $h_{t}=1$
indicates that the transmission is successful, and $h_{t}=0$, otherwise.
We define the success probability as $\Pr\{h_{t}=1\}=p_{s}$ and the
failure probability as $\Pr\{h_{t}=0\}=p_{f}=1-p_{s}$. Upon receiving
the update packet, the destination feeds back a single-bit acknowledgement,
which is assumed to be instant and error-free. If the transmission
is failed and the sensor decides to transmit in the next slot, it
would generate and transmit a new status update rather than retransmit
the failed update. This is because, with the same success probability,
retransmitting the failed out-of-date status update leads to a larger
age. 

\subsection{Freshness Metric}

We assume that a status update is generated and transmitted at the
beginning of a slot and it will be received by the end of the slot
if the transmission succeeds. AoI, which is usually used to quantify
the information freshness, is defined as the time elapsed since the
generation of the latest status update received by the destination.
Suppose that the update $i$ is generated and delivered at the time
instants $g_{i}$ and $d_{i}$, respectively. Let $U(t)$ denote the
time at which the latest status update successfully received by the
destination was generated, i.e., $U(t)=\max\{g_{i}\mid d_{i}\leq t\}$.
The AoI at the beginning of slot t is then given by
\begin{equation}
\delta_{t}=t-U(t).
\end{equation}

Different from AoI, our proposed metric, AoCI, not only captures the
time lag of the received update at the destination, but also incorporates
the variation of the information content of the update. In particular,
the AoCI decreases only when the content of the newly received update
is different from the previous one, and increases otherwise. Let
$n(t)=\max\{i|d_{i}\leq t\}$ be the index of the latest update received
by the destination at the beginning of slot $t$ and $m(t)=\max\{j|Y_{j}\neq Y_{n(t)},d_{j}\leq d_{n(t)}\}$
be the index of the most recently update that has different content
from the latest received update. $Y_{j}$ denotes the information
content of update $j$, which is equal to the state of the physical
process in the slot when update $j$ was generated. Then, we can define
the AoCI at the beginning of slot $t$ as
\begin{equation}
\Delta_{t}=t-U'(t),\label{eq:AoCI}
\end{equation}
where $U'(t)=\min\{g_{k}|d_{m(t)}<d_{k}\leq d_{n(t)}\}$ represents
the generation time of the next successfully received update packet
after $m(t)$. It is worth noting that all the successfully received
update packets after $m(t)$ has the same content with the latest
received one.

\begin{figure}[t]
\centering

\includegraphics[width=0.5\textwidth]{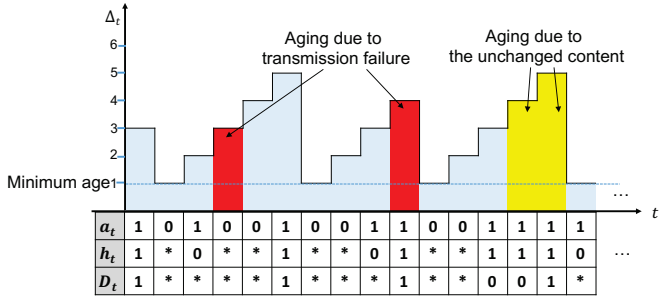}\caption{\label{fig:AOI figure}An illustration of the AoCI in a time-slotted
status update system, where $*$ is used to represent the irrelevant
values.}
\end{figure}

Let $D_{t}\in\{0,1\}$ denote whether the content of a newly received
update is different from that of the previously received one. If $D_{t}=1$,
then the newly received update has different content. Otherwise, it
has the same content. We define $p_{r}=\Pr(D_{t}=1)=\Pr(Y_{n(t)}=Y_{n(t)-1})$.
Note that $Y_{n(t)}=X_{U(t)}$, we have $p_{r}=\Pr(X_{U(t)}=X_{U(t)-\delta})$,
which is the return probability that a state of the physical process
does not change after $\delta$ steps. According to (\ref{eq:AoCI}),
if a new status update generated by the sensor is successfully received
by the destination (i.e., $a_{t}=1,h_{t}=1$) and it contains different
content from the previously received update (i.e., $D_{t}=1$), then
the AoCI decreases to one; otherwise, the AoCI increases by one. Then,
the dynamics of the AoCI can be given by
\begin{equation}
\Delta_{t+1}=\begin{cases}
1 & a_{t}=1,h_{t}=1,D_{t}=1;\\
\Delta_{t}+1, & \text{otherwise}.
\end{cases}\label{eq:Dynamic}
\end{equation}
For ease of exposition, we use Fig. \ref{fig:AOI figure} to illustrate
the evolution of AoCI over time. 

\subsection{Problem Formulation}

The objective of this paper is to find an update policy $\pi=(a_{0},a_{1},\ldots)$
that minimizes the total average cost, which is the weighted sum of
the AoCI and the update cost. By defining $\Pi$ as a set of stationary
policies, our problem can be formulated as follows:
\begin{equation}
\min_{\pi\in\Pi}\limsup_{T\rightarrow\infty}\frac{1}{T}\stackrel[t=0]{T}{\sum}\mathbb{E}[\Delta_{t}+\omega a_{t}C_{u}|s_{0}],\label{eq:Problem}
\end{equation}
where $\omega$ is a weighting factor and is used to reflect the levels
of importance and $s_{0}$ is the initial state.

\section{Updating Policy Design \label{sec:Updating-Policy-Design}}

\subsection{MDP Characterization}

The optimization problem in (\ref{eq:Problem}) can be cast into an
infinite horizon average cost Markov decision process $(\mathcal{S},\mathcal{A},\Pr(\cdot|\cdot,\cdot),C(\cdot,\cdot))$,
where each item is explained as follows:
\begin{itemize}
\item States: The state of the MDP in time slot $t$ is defined to be the
tuple of AoCI and AoI, i.e., $s_{t}\triangleq\left(\Delta_{t},\delta_{t}\right)$,
which can take any value in $\mathbb{Z}^{+}\times\mathbb{Z}^{+}$.
Therefore, the state space $\mathcal{S}$ is countable and infinite.
\item Actions: The action in time slot $t$ is $a_{t}$ and the action set
$\mathcal{A}=\{0,1\}$ is finite and countable. 
\item Transition Probability: Let $\Pr(s_{t+1}|s_{t},a_{t})$ denote the
transition probability that state transits from $s_{t}$ to $s_{t+1}$
in the next slot by taking action $a_{t}$ in slot $t$. Since the
failure of the packet transmission and the content change of the received
updates are independent, according to the AoCI evolution dynamics
(\ref{eq:Dynamic}), the transition probability can be written as
\begin{equation}
\begin{cases}
\Pr(s_{t+1}=(\Delta+1,\delta+1)|s_{t}=(\Delta,\delta),a_{t}=0)=1,\\
\Pr(s_{t+1}=(\Delta+1,\delta+1)|s_{t}=(\Delta,\delta),a_{t}=1)=p_{f},\\
\Pr(s_{t+1}=(\Delta+1,1)|s_{t}=(\Delta,\delta),a_{t}=1)=p_{s}p_{r}(\delta),\\
\Pr(s_{t+1}=(1,1)|s_{t}=(\Delta,\delta),a_{t}=1)=p_{s}(1-p_{r}(\delta)),
\end{cases}\label{eq:state-transition}
\end{equation}
and $\Pr(s_{t+1}|s_{t},a_{t})=0$ otherwise. 
\item Cost: Let $C(s_{t},a_{t})$ denote the instantaneous cost at state
$s_{t}$ given action $a_{t}$, which is given by $C(s_{t},a_{t})=\Delta_{t}+\omega a_{t}C_{u}$.
\end{itemize}

The optimal policy $\pi^{*}$ to minimize the total average cost can
be obtained by solving the following Bellman equation \cite{dimitrip.bertsekasDynamicProgrammingOptimal2007}:
\begin{equation}
\theta+V(s)=\min_{a\in\{0,1\}}\left\{ C(s,a)+\sum_{s'\in\mathcal{S}}\Pr(s'|s,a)V(s')\right\} ,\forall s\in\mathcal{S},\label{eq:Bellman}
\end{equation}
where $\theta$ is the optimal value to (\ref{eq:Problem})  and
$V(s)$ is the value function which is a mapping from $s$ to real
values. Moreover, for any $s\in\mathcal{S}$, the optimal policy can
be given by
\begin{equation}
\pi^{*}(s)=\arg\min_{a\in\{0,1\}}\left\{ C(s,a)+\sum_{s'\in\mathcal{S}}\Pr(s'|s,a)V(s')\right\} .\label{eq:optimal-policy}
\end{equation}
It can be seen from (\ref{eq:optimal-policy}) that the optimal policy
$\pi^{*}$ depends on the value function $V(\cdot)$, for which there
is no closed-form solution in general \cite{dimitrip.bertsekasDynamicProgrammingOptimal2007}.
In the literature, various numerical algorithms, such as value iteration
and policy iteration, have therefore been proposed. However, these
methods are usually computationally demanding due to the curse of
dimensionality and few insights for the optimal policy can be leveraged.
Therefore, we study the structural properties of the optimal updating
policy in the sequel.

\subsection{Structural Analysis and Optimal Policy }

We consider a special case that $p_{c}=1/2$. In this case, the return
probability $p_{r}(\delta)=1/2$ for all $\delta$. In other word,
$p_{r}$ is irrespective of $\delta$. Hence, we can simplify the
states of the MDP. In particular, the state in slot $t$ reduces to
the AoCI, i.e., $s_{t}=\Delta_{t}$, and the state transition probability
in (\ref{eq:state-transition}) can be simplified as 
\begin{equation}
\begin{cases}
\Pr(s_{t+1}=\Delta+1|s_{t}=\Delta,a_{t}=0)=1,\\
\Pr(s_{t+1}=\Delta+1|s_{t}=\Delta,a_{t}=1)=p_{f}+p_{s}p_{r},\\
\Pr(s_{t+1}=1|s_{t}=\Delta,a_{t}=1)=p_{s}(1-p_{r}),
\end{cases}
\end{equation}
and $\Pr(s_{t+1}|s_{t},a_{t})=0$ otherwise. Based on the simplified
state space and transition probability, we present the monotonicity
property of $V(s)$ in the following lemma. 
\begin{lem}
\label{lem:lemma2}The value function V(s) is a non-decreasing function
for $s\in\mathcal{S}$. 
\end{lem}
\begin{IEEEproof}
See Appendix \ref{subsec:Proof-of-Lemma 2}.
\end{IEEEproof}
Then, we provide results on the structure of the optimal updating
policy in the following theorem.
\begin{thm}
\label{thm:theorem1}For $s\in\mathcal{S}$, the optimal policy has
a switching structure, that is if $\pi^{*}(s_{1})=1$, then $\pi^{*}(s_{2})=1$
for all $s_{2}\geq s_{1}$.
\end{thm}
\begin{IEEEproof}
See Appendix \ref{subsec:Proof-of-Theorem-1}.\emph{}
\end{IEEEproof}
According to Theorem \ref{thm:theorem1}, the optimal policy can be
represented as a threshold policy, which is given by
\begin{equation}
\pi^{*}(s)=\begin{cases}
1, & \text{if }s\ge\Omega^{*},\\
0, & \text{otherwise},
\end{cases}\label{eq:Threshold}
\end{equation}
where $\Omega^{*}$ is the threshold at which the switching occurs.
Thanks to the simplifications in the special case, we are able to
derive the closed-form of $\Omega^{*}$.
\begin{thm}
\label{thm:closed-form} The optimal threshold $\Omega^{*}$ of the
threshold policy is given by
\begin{equation}
\Omega^{*}=\frac{\sqrt{p_{z}+2\omega C_{u}(1-p_{z})}-p_{z}}{1-p_{z}},
\end{equation}
where $p_{z}=p_{f}+p_{s}p_{r}$.
\end{thm}
\begin{IEEEproof}
See Appendix \ref{subsec:Proof-of-closed-form}.
\end{IEEEproof}
If $\Omega^{*}$ is an integer, the optimal policy is shown in (\ref{eq:Threshold}).
Otherwise, the optimal policy is given by
\begin{equation}
\pi^{*}(s)=\begin{cases}
1, & \text{if }s\ge\left\lceil \Omega^{*}\right\rceil ,\\
{\bf 1}_{(x\leq\mu)}, & \text{if }s=\left\lfloor \Omega^{*}\right\rfloor ,\\
0, & \text{if }s<\left\lfloor \Omega^{*}\right\rfloor ,
\end{cases}
\end{equation}
where $\bm{1}_{(\cdot)}$ is an indicator function, $x\in[0,1]$ is
a uniform random variable, and $\mu=\frac{\left\lceil \Omega^{*}\right\rceil -\Omega^{*}}{\left\lceil \Omega^{*}\right\rceil -\left\lfloor \Omega^{*}\right\rfloor }$.
Specifically, $\pi^{*}(\left\lfloor \Omega^{*}\right\rfloor )=1$
with probability $\mu$ and $\pi^{*}(\left\lfloor \Omega^{*}\right\rfloor )=0$
with probability $1-\mu$.

\section{Simulation Results}

In this section, we present the simulation results of the optimal
updating policy to investigate the effects of system parameters and
compare the optimal updating policy with zero-wait policy.

\begin{figure}[tp]
\centering

\includegraphics[width=0.5\textwidth]{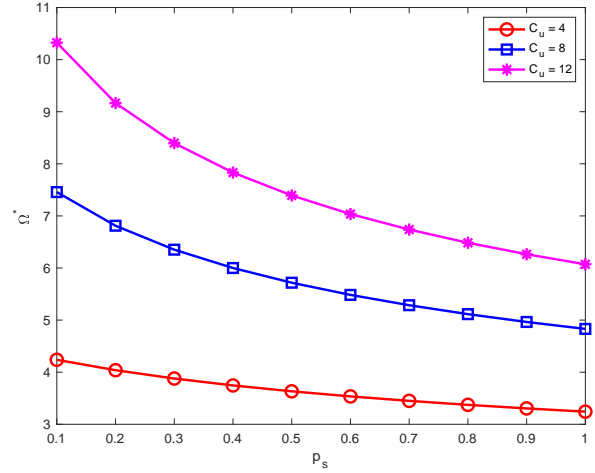}\caption{\label{fig:Optimal threshold}The optimal threshold for different
values of $p_{s}$ ($p_{c}=0.5$ and $\omega=1$).}
\end{figure}

Fig. \ref{fig:Optimal threshold} shows the optimal threshold of the
optimal updating policy with respect to $p_{s}$ for different $C_{u}$.
It can be seen that the larger the cost, the larger the threshold
is. This is evident from Theorem \ref{thm:closed-form}. We can observe
that the smaller the $p_{s}$, the larger the threshold is. This is
because, when $p_{s}$ is small, the sensor has to sample and transmit
multiple times until the destination successfully receives an update
packet. Therefore, it is efficient to update the status only when
the AoCI is large. 

\begin{figure}[tp]
\centering

\includegraphics[width=0.5\textwidth]{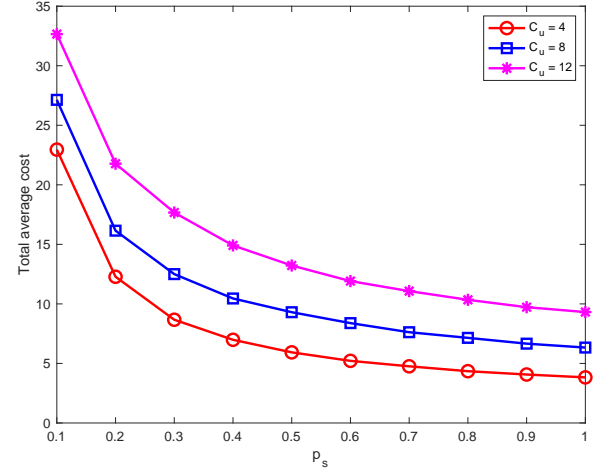}\caption{\label{fig:Impact of Energy cost}Effect of $p_{s}$ on the total
average cost for different values of $C_{u}$ ($p_{c}=0.5$ and $\omega=1$).}
\end{figure}

Fig. \ref{fig:Impact of Energy cost} illustrates the total average
cost of the optimal policy with respect to $p_{s}$ for different
$C_{u}$. The effect of $p_{s}$ on the performance can be seen immediately:
the larger the $p_{s}$, the smaller the total average cost is. As
$p_{s}$ increases, the transmission of an update is much easier to
be successful, and hence the average AoCI and the average update cost
are both reduced. Moreover, larger $C_{u}$ results in an increase
in the total average cost as expected, and the gap between the total
average cost for different $C_{u}$ values is almost constant with
respect to $p_{s}$.

\begin{figure}[tp]
\centering

\includegraphics[width=0.5\textwidth]{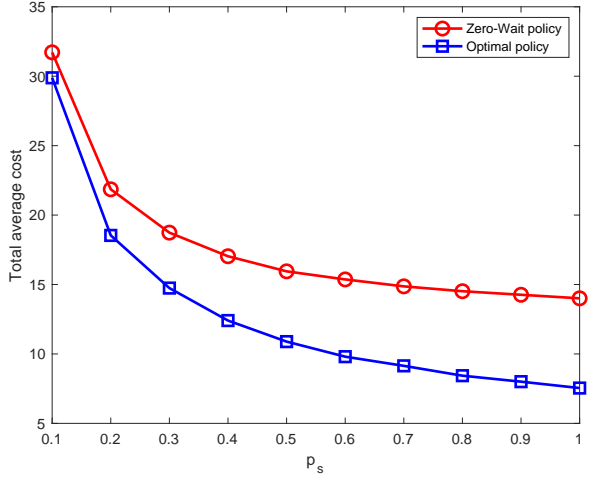}\caption{\label{fig:Comparision bewteen optimal policy and zero-wait policy-1}Comparison
between the optimal policy and zero-wait policy in terms of the total
average cost ($p_{c}=0.5$, $C_{u}=12$ and $\omega=1$).}
\end{figure}

\begin{figure}[tp]
\centering

\includegraphics[width=0.5\textwidth]{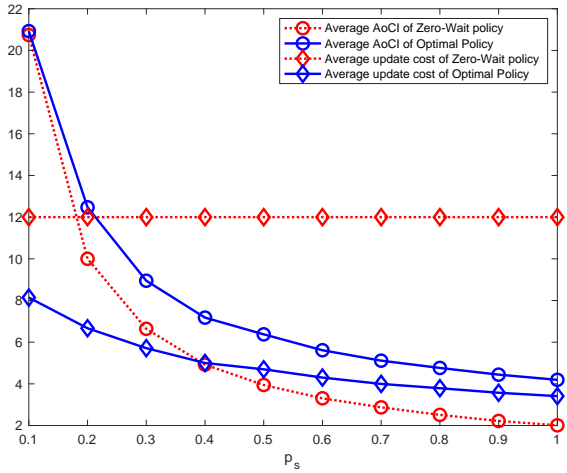}\caption{\label{fig:Comparision of AoCI and cost}Comparison between the optimal
policy and zero-wait policy in terms of the average AoCI and the average
update cost ($p_{c}=0.5$, $C_{u}=12$ and $\omega=1$).}
\end{figure}

In Fig. \ref{fig:Comparision bewteen optimal policy and zero-wait policy-1},
we compare the total average cost of the optimal policy and the zero-wait
baseline policy. In the zero-wait policy, the sensor samples and transmits
the status update in each time slot. We can see that the optimal policy
is superior to the zero-wait policy and the reduction of the total
average cost increases with increasing $p_{s}$. This is due to the
fact, as shown in Fig. \ref{fig:Comparision of AoCI and cost}, that
the zero-wait policy achieves a smaller AoCI but suffers from a constant
update cost, while the optimal policy can strike a balance between
the AoCI and the update cost. In particular, compared with the zero-wait
policy, the optimal policy has a larger AoCI because the sensor remains
idle until the AoCI is larger than a threshold. However, its update
cost decreases as $p_{s}$ grows and hence the optimal policy is more
cost-efficient. 


\section{Conclusion}

In this paper, we have proposed a new freshness metric that addresses
the ignorance of information content in the conventional AoI. Named
as the age of changed information, this new metric not only measures
the freshness by the passage of time but also captures the information
content of the updates at the destination. We have studied the updating
policy in the status update system by taking both the AoCI and the
update cost into consideration and formulated the updating problem
as an infinite horizon average cost MDP. We have shown that the optimal
updating policy in a special case is of threshold type, which reveals
an intrinsic tradeoff between the average AoCI and the update cost.
Simulation results have shown the effects of the unreliable channel
on the total average cost. Through the comparison between the threshold
policy and the zero-wait policy, the threshold policy is shown to
yield significant performance gain in terms of the total average cost
compared to a zero-wait policy. Future work will address some extensions
such as modeling the physical process with a more general Markov chain
model and incorporating time-correlated channel statistics.

\appendix{}

\subsection{Proof of Lemma \ref{lem:lemma2}\label{subsec:Proof-of-Lemma 2}}

Based on the value iteration algorithm (VIA) \cite{dimitrip.bertsekasDynamicProgrammingOptimal2007},
we use mathematical induction to prove Lemma \ref{lem:lemma2}. For
each state $s$, let $V_{k}(s)$ be the value function at iteration
$k$. In VIA, the value function can be updated as follows:
\begin{equation}
V_{k+1}(s)=\min_{a}\left\{ C(s,a)+\sum_{s'\in\mathcal{S}}\Pr(s'|s,a)V_{k}(s')\right\} ,\forall s\in\mathcal{S}.
\end{equation}
Under any initialization of the initial value $V_{0}(s)$, the sequence
$\{V_{k}(s)\}$ converges to the value function in the Bellman equation
(\ref{eq:Bellman}) \cite{dimitrip.bertsekasDynamicProgrammingOptimal2007},
i.e.,
\begin{equation}
\lim_{k\rightarrow\infty}V_{k}(s)=V(s),\forall s\in\mathcal{S}.\label{eq:lim-value-function}
\end{equation}
Therefore, the monotonicity of $V(s)$ in $\mathcal{S}$ can be guaranteed
by proving that for any $s_{1},s_{2}\in\mathcal{S}$, such that $s_{1}\leq s_{2}$,
\begin{equation}
V_{k}(s_{1})\leq V_{k}(s_{2}),\quad k=0,1,\ldots\label{eq:monotonicity}
\end{equation}

Then, we prove (\ref{eq:monotonicity}) via mathematical induction.
Without loss of generality, we initialize $V_{0}(s)=0$ for all $s\in\mathcal{S}$.
Thus, (\ref{eq:monotonicity}) holds for $k=0$. Next, we assume that
(\ref{eq:monotonicity}) holds up till $k>0$ and we examine whether
it holds for $k+1$. Let $Q_{k}(s,a)$ denote the state-action value
function at iteration $k$, which is defined as
\begin{equation}
Q_{k}(s,a)=C(s,a)+\sum_{s'\in\mathcal{S}}\Pr(s'|s,a)V_{k}(s'),
\end{equation}
for all $s\in\mathcal{S}$ and $a\in\mathcal{A}$. Then, the value
function at iteration $k+1$ can be represented as 
\begin{equation}
V_{k+1}(s)=\min\limits _{a\in\{0,1\}}Q_{k}(s,a).
\end{equation}

When $a=0$, we have $Q_{k}(s_{1},0)=s_{1}+\mathop{V_{k}(s_{1}+1)}$
and $Q_{k}(s_{2},0)=s_{2}+\mathop{V_{k}(s_{2}+1)}$. Since $s_{1}\leq s_{2}$
and $V_{k}(s_{1})\leq V_{k}(s_{2})$, we can easily see that $Q_{k}(s_{1},0)\leq Q_{k}(s_{2},0)$.

When $a=1$, we have 
\begin{align*}
Q_{k}(s_{1},1)= & s_{1}+\omega C_{u}\\
 & +(p_{f}+p_{s}p_{r})\mathop{V_{k}(s_{1}+1)}+p_{s}(1-p_{r})V_{k}(1)
\end{align*}
 and 
\begin{align*}
Q_{k}(s_{2},1)= & s_{2}+\omega C_{u}\\
 & +(p_{f}+p_{s}p_{r})\mathop{V_{k}(s_{2}+1)}+p_{s}(1-p_{r})V_{k}(1).
\end{align*}
 Bearing in mind that $V_{k}(s_{1})\leq V_{k}(s_{2})$, we can also
verify that $Q_{k}(s_{1},1)\leq Q_{k}(s_{2},1)$.

Altogether, we can assert that $V_{k+1}(s_{1})\leq V_{k+1}(s_{2})$
for any $k$. By taking limits on both sides of (\ref{eq:monotonicity})
and by (\ref{eq:lim-value-function}), we complete the proof of Lemma
\ref{lem:lemma2}. 

\subsection{Proof of Theorem \ref{thm:theorem1} \label{subsec:Proof-of-Theorem-1}}

Let $Q(s,a)$ denote the state-action value function, i.e., 
\begin{equation}
Q(s,a)=s+\omega aC_{u}+\mathop{\sum_{s'\in\mathcal{S}}\Pr(s'|s,a)V(s')}.
\end{equation}
 The optimal policy can be expressed as 
\begin{equation}
\pi^{*}(s)=\arg\min\limits _{a\in\{0,1\}}Q(s,a).
\end{equation}
 Suppose $\pi^{*}(s_{1})=1$, we have $Q(s_{1},0)-Q(s_{1},1)\geq0$.
Therefore, the optimal updating policy has a switching structure if
$Q(s,a)$ has a sub-modular structure, that is, 
\begin{equation}
Q(s_{1},0)-Q(s_{1},1)\leq Q(s_{2},0)-Q(s_{2},1),\label{eq:sub-modular}
\end{equation}
for any $s_{1},s_{2}\in\mathcal{S}$ and $s_{1}\leq s_{2}$.

According to the definition of $Q(s,a)$, we have 
\begin{align*}
 & Q(s_{1},0)-Q(s_{1},1)\\
= & p_{s}(1-p_{r})(V(s_{1}+1)-V(1))-\omega C_{u}
\end{align*}
 and 
\begin{align*}
 & Q(s_{2},0)-Q(s_{2},1)\\
= & p_{s}(1-p_{r})(V(s_{2}+1)-V(1))-\omega C_{u}.
\end{align*}
 Since $V\left(s_{1}+1\right)\leq V(s_{2}+1)$, it is easy to see
that (\ref{eq:sub-modular}) holds. Along with $Q(s_{1},0)-Q(s_{1},1)\geq0$,
we complete the proof of Theorem \ref{thm:theorem1}.

\subsection{Proof of Theorem\ref{thm:closed-form} \label{subsec:Proof-of-closed-form}}

\begin{figure}[tp]
\centering

\includegraphics[width=0.5\textwidth]{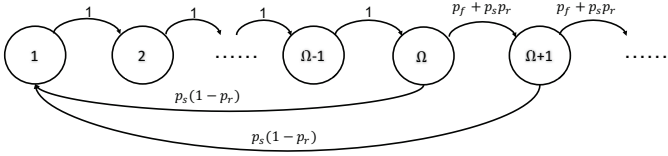}\caption{\label{fig:StateTransition} The states transitions under a threshold
policy.}
\end{figure}
For any threshold policy with the threshold of $\Omega$, the MDP
can be modeled through a Discrete Time Markov Chain (DTMC) with the
same states, which is illustrated in Fig. \ref{fig:StateTransition}.
Let $\varphi_{s}$ denote the steady state probability of state $s$.
According to Fig. \ref{fig:StateTransition}, we have 
\begin{equation}
\varphi_{s}=\begin{cases}
\varphi_{1}, & \text{if }s\le\Omega,\\
\varphi_{1}p_{z}^{s-\Omega}, & \text{otherwise},
\end{cases}
\end{equation}
where $p_{z}=p_{f}+p_{s}p_{r}$. Along with $\stackrel[i=1]{\infty}{\sum}\varphi_{i}=1$,
we can derive $\varphi_{s}$ in closed-form as follows:
\begin{equation}
\varphi_{s}=\begin{cases}
\frac{1-p_{z}}{\Omega(1-p_{z})+p_{z}}, & \text{if }s\le\Omega,\\
\frac{(1-p_{z})p_{z}^{s-\Omega}}{\Omega(1-p_{z})+p_{z}}, & \text{otherwise}.
\end{cases}
\end{equation}

Then, the expected cost under the threshold policy can be computed
as:
\begin{align}
J_{\Omega} & =\stackrel[s=1]{\infty}{\sum}\varphi_{s}(s+\omega C_{u}{\bf 1}_{(s\geq\Omega)})\nonumber \\
 & =\sum_{s=1}^{\Omega-1}\varphi_{s}s+\sum_{s=\Omega}^{\infty}\varphi_{s}(s+\omega C_{u})\nonumber \\
 & =\frac{1-p_{z}}{\Omega(1-p_{z})+p_{z}}\left(\frac{\Omega^{2}-\Omega}{2}+\frac{\Omega+\omega C_{u}}{1-p_{z}}+\frac{p_{z}}{(1-p_{z})^{2}}\right).\label{eq:Expected cost function}
\end{align}
Since $J_{\Omega}$ is a convex function of $\Omega$ by (\ref{eq:Expected cost function}),
the optimal threshold can be obtained by setting the derivative $\partial J_{\Omega}/\partial\Omega$
to zero. Specifically, 
\begin{equation}
\Omega^{*}=\frac{\sqrt{p_{z}+2\omega C_{u}(1-p_{z})}-p_{z}}{1-p_{z}},
\end{equation}
which concludes our proof.

\bibliographystyle{IEEEtran}
\bibliography{AoI}

\end{document}